\documentclass[preprint,prc,aps,showpacs]{revtex4}
\usepackage{epsfig}
\usepackage{amsmath}
\usepackage[hyperref]{hyperref}

\newcommand{\Pomeron}{I\!\!P}

\begin{document}

\title{Leading twist nuclear shadowing and suppression of hard coherent 
diffraction in proton-nucleus scattering}

\author{V.~Guzey}
\affiliation{Institut f{\"u}r Theoretische Physik II,
Ruhr-Universit{\"a}t Bochum,
  D-44780 Bochum, Germany}
\email[]{vadim.guzey@tp2.ruhr-uni-bochum.de}

\author{M.~Strikman}
\affiliation{Department of Physics, Pennsylvania State University,
University Park, PA 16802, USA}
 \email[]{strikman@phys.psu.edu}

\preprint{RUB-TPII-7/2005}
\pacs{24.85.+p,25.40.Ep,13.87.-a}

\begin{abstract}

We use the  Glauber-Gribov multiple scattering formalism and the 
theory of leading twist nuclear shadowing to 
develop a method for the calculation of leading twist hard coherent 
diffraction in hadron-nucleus  processes. We demonstrate that soft multiple 
rescatterings lead to the factorization breaking of hard diffraction in proton-nucleus  
scattering, which is larger than in hadron-nucleon scattering.
At the LHC and RHIC kinematics, we compare 
the hard diffractive to e.m.~mechanisms of
hard coherent production of two jets in proton-nucleus scattering.
We study the $x_{\Pomeron}$, $\beta$ and $A$-dependence
of the ratio of the dijet production 
cross sections due to the two effects, $R$.
We demonstrate that in proton-heavy nucleus hard coherent diffraction at the LHC,
$R$ is small, which offers 
 a clean method to study  hard photon-proton scattering at
 the energies exceeding the HERA energies by the factor of ten.
On the other hand, the use of lighter nuclei, such as $^{40}$Ca, will allow to study the
screened nuclear diffractive parton distribution. Moreover, a comparison
of the dijet diffractive production to the heavy-quark-jet diffractive production
will estimate the screened nuclear diffractive gluon PDF, which
will be measured in nucleus-nucleus ultraperipheral collisions at the LHC.

\end{abstract}

\maketitle

\section{Introduction}
\label{sec:intro}

In hadron-hadron scattering at high energies, diffractive processes
are characterized by the rapid $t$-dependence and 
by the absence of detected particles in a certain region
of the final phase space or, in other words, by the presence of the rapidity gap.
When a hard scale is present in diffractive scattering, such processes are called
hard diffractive.
The phenomenon of hard diffraction in proton-antiproton scattering
was first discovered in the $p\, \bar{p} \to p \,X$ reaction at the 
the SPC collider at CERN, 
when it was observed that the diffractive 
final state $X$ with the invariant mass in the range 105 to 190 GeV
contained 
jets with the transverse energy between 5
and 13 GeV 
(the hard scale is given by the jet transverse 
momentum)~\cite{Bonino:1988ae}. 
Later, hard diffraction in proton-antiproton scattering
was studied at the Tevatron in dijet, $W$, $b$-quark and $J/\Psi$ production, see~\cite{Goulianos:2004as} for a review.
The cross section of each hard diffractive channel constitutes
approximately 1\% of the contribution of
the corresponding channel to the inclusive $p\, \bar{p}$ cross section.

Turning to electron-proton deep inelastic scattering (DIS), it was one of HERA suprises
to observe that hard diffractive events characterized by a large rapidity gap
between the virtual photon and the proton fragmentation regions
constitute about 10\% of the total rate of events~\cite{Abramowicz:1998ii}.
 In DIS, the hard scale is
provided by the virtuality of the photon, $Q^2$.

In the theoretical treatment of 
soft
diffractive processes, the key role is played by the
concept of the "Pomeron", the Regge trajectory with vacuum quantum numbers, which
provides the diffractive exchange and
determines the high-energy behavior of elastic and diffractive scattering
amplitudes. In the context of hard diffraction, the notion of
Pomeron appears as follows.
The QCD factorization theorem for hard diffraction in 
DIS~\cite{Collins:1997sr} enables one to introduce universal
diffractive parton distribution
functions (PDFs), which can relate such processes as inclusive diffraction,
dijet diffractive production, $D^{\ast}$-meson diffractive production, etc.
Making an assumption that the diffractive PDFs can be factorized into the product of
two terms, representing the Pomeron flux and the Pomeron parton distributions, one can
effectively study the parton content of the Pomeron, similarly to
the parton content of the nucleon in inclusive 
DIS~\cite{Breitweg:1998gc,Adloff:1997sc,Adloff:2000qi,unknown:2006hy,:2006hx}.

A comparison of hard diffraction in proton-antiproton scattering
at the Tevatron to hard diffraction in electron-proton scattering
at HERA indicates the breakdown of the QCD factorization: The use
of diffractive PDFs extracted from the HERA measurements
significantly 
overestimates
the rates of hard diffraction at 
the Tevatron~\cite{Goulianos:2004as,Goulianos:1995wy}.
This can be explained by the absorptive effects
associated with multi-Pomeron exchanges, which
 make the gap survival very unlikely in the case of hadronic  
collisions~\cite{Kaidalov:2001iz}, or by the gradual onset
of the so-called black disk regime for the proton-proton collisions
at small impact parameters~\cite{Frankfurt:2004ti}.

Turning to diffraction in hadron-nucleus and lepton-nucleus scattering,
the situation can be briefly summarized as follows.
Soft coherent (without the nuclear break-up) diffraction in hadron-nucleus
scattering at high energies can be successfully described within
the framework of the Glauber-Gribov approach by taking into account cross section
(color) fluctuations in the hadronic 
projectile~\cite{Frankfurt:1993qi,Strikman:1995jf,Frankfurt:2000ty,Guzey:2005tk},
see also Sect.~\ref{sec:formula}.

In DIS on nuclear targets, nuclear diffractive PDFs at small values of Bjorken $x$
can be expressed in terms of the nucleon diffractive PDFs, which are known from the
HERA studies~\cite{Frankfurt:2003gx}.
This approach to nuclear diffractive PDFs and to usual nuclear parton distribution
functions is based on the Gribov's connection between the nuclear shadowing
correction and the elementary diffractive cross section, 
the QCD factorization theorem for hard diffraction in DIS~\cite{Collins:1997sr} and
the QCD analysis of HERA data on hard diffraction~\cite{Breitweg:1998gc,Adloff:1997sc,Adloff:2000qi,unknown:2006hy,:2006hx},
see~\cite{Frankfurt:2003zd} and references therein and also Sect.~\ref{sec:formula}.

In this work, we study hard diffraction in 
proton-nucleus collisions.
As an example, we derive the expression for the cross section of the 
hard coherent diffractive production of two jets in proton-nucleus
scattering. We analyze
the suppression of hard
diffraction in proton-nucleus scattering compared to hard diffraction in
proton-proton scattering. A numerical analysis of the corresponding
 suppression factor
enables us to quantify
the QCD factorization 
breaking in hard proton-nucleus diffraction due to the soft
screening (absorption).

The paper consists of two parts.
 In Sect.~\ref{sec:formula}, we derive a general 
expression for the nuclear
 modifications of hard diffraction in proton-nucleus scattering. 
We demonstrate that because of multiple soft rescatterings,
 hard diffractive processes such as production of two jets, heavy flavors,
 etc.~are
 suppressed at the LHC
and RHIC 
energies stronger than soft inelastic diffraction,
 which in turn is expected to be strongly
 suppressed~\cite{Frankfurt:1993qi,Frankfurt:2000ty,Guzey:2005tk}.

In Sect.~\ref{sec:ultraperipheral}, we compare the contribution of hard 
coherent proton-nucleus diffraction into two jets (including heavy-quark jets) to the 
e.m.~contribution, when the final state containing two hard jets is produced
by the coherent nuclear Coulomb field. We demonstrate that  the 
e.m.~contribution dominates proton-heavy nucleus (such as $^{208}$Pb) scattering at the LHC, which
provides essentially a background-free method to study very high-energy 
$\gamma \,p$ 
scattering at the LHC through ultraperipheral proton-nucleus scattering.
We show that using lighter nuclei, which do not produce such a
 strong flux of equivalent 
photons as $^{208}$Pb, one can study screened nuclear diffractive PDFs. In this case,
a comparison
of the dijet diffractive production to the heavy-quark-jet diffractive production
will measure the nuclear screened diffractive gluon PDF.
 The latter can be compared to the nuclear
diffractive PDFs, which will be measured in nucleus-nucleus ultraperipheral
collisions at the LHC.
The conclusion about the dominance of the hard diffractive mechanism over the e.m.~one,
 when light nuclei are used, also holds in the RHIC kinematics.

 Our results are also valid for the diffraction in  resolved photon-nucleus interactions.
 Since in this case several other effects are also important, we will discuss
 hard diffraction in $\gamma\, A$ interactions in a separate publication.

\section{Suppression factor for hard proton-nucleus diffraction}
\label{sec:formula}

The derivation of the expression for the amplitude of
hard diffraction in hadron-nucleus
scattering combines features of soft coherent diffraction in hadron-nucleus
scattering and 
hard coherent diffraction in DIS on
nuclear targets. Therefore, we shall briefly review coherent soft and hard 
diffraction below.

\subsection{Soft coherent proton-nucleus diffraction}

At high energies, the cross section of soft coherent hadron-nucleus diffraction
(diffraction dissociation),
$\sigma_{DD}^{hA}$,
can be economically and reliably calculated using the Glauber-Gribov multiple
scattering formalism~\cite{Glauber:1955qq,Glauber:1970jm,Gribov:1968jf} generalized to include the so-called cross section 
(color) fluctuations
in the projectile~\cite{Frankfurt:1993qi,Strikman:1995jf,Frankfurt:2000ty,Guzey:2005tk}
\begin{equation}
\sigma_{DD}^{hA}=\int d^2 \, b \, \left(
  \int d \sigma P(\sigma)\left|\Gamma_A(b,\sigma) \right|^2 -
 \left| \int d \sigma P(\sigma)\Gamma_A(b,\sigma)\right|^2 \right) \,.
\label{scf6}
\end{equation}
In Eq.~(\ref{scf6}),
$b$ is the impact parameter (the two-dimensional vector
connecting the trajectory of the projectile with the center of the target 
nucleus);
$P(\sigma)$ is the probability to find in the projectile a hadronic
configuration, which interacts with target nucleons with the cross section
$\sigma$; $\Gamma_A$ is the projectile-nucleus scattering amplitude
in the impact parameter space~\cite{Glauber:1970jm}
\begin{equation}
\Gamma_A(b,\sigma)=1-\exp \left(-\frac{1}{2} \, \sigma \, T_A(b) \right)
 \,,
 \label{scf5}
\end{equation}
where $T_A(b)=\int dz \,\rho_A(b,z)$ and 
$\rho_A$ is the nucleon density normalized to 
the number of the nucleons $A$.
The energy-dependence of $\sigma_{DD}^{hA}$ is determined by the 
energy-dependence of $P(\sigma)$, which is implied~\cite{Guzey:2005tk}.
Note that in Eq.~(\ref{scf5}),  we neglected the slope of the elementary
hadron-nucleon scattering amplitude compared to the nuclear size and
we assumed that the elementary scattering amplitude is purely imaginary, which is a good approximation at the
high energies that we consider in this work.

The function $P(\sigma)$ describes the probability that the incoming hadron 
interacts with target nucleons with a given cross section $\sigma$.
In other words, $P(\sigma)$ describes cross section fluctuations
in the energetic projectile. The notion of $P(\sigma)$ 
is introduced in order to have a compact phenomenological description
of soft coherent diffraction in hadron-nucleon and
hadron-nucleus scattering.
As follows from Eq.~(\ref{scf6}),
ignoring cross section fluctuation, i.e.~setting 
$P(\sigma) \propto \delta(\sigma-\sigma_{{\rm tot}})$, would result in 
the unacceptable result that $\sigma_{DD}^{hA}=0$.

The formalism of cross section fluctuations is based on the
simple picture of  diffractive dissociation in the laboratory
reference frame developed by
Feinberg and Pomeranchuk~\cite{Feinberg} and by
Good and Walker~\cite{GW}. In this picture, the incoming hadron
is represented by a coherent superposition of eigenstates of the
scattering operator. Since different eigenstates correspond to
different $\sigma$, the scattered state is in general different from
the incoming state, but it has the same quantum numbers.
This corresponds to the process of diffractive dissociation.

One should note that the formalism of cross section fluctuations 
implicitly uses the assumption of the completness of the scattering
eigenstates and, hence, it is 
applicable only at $t \approx 0$. At $t \neq 0$, the diffractive
final state can be produced as a result of some effective interaction or
as a result of hard parton scattering (for sufficiently large $t$), which 
have nothing to do with the cross section fluctuations in the projectile.

The function $P(\sigma)$ is different for different projectiles
(protons, pions, photons). For the proton, $P(\sigma)$ has a narrow
dispersion around $\sigma=\sigma_{{\rm tot}}$, where $\sigma_{{\rm tot}}$
is the total proton-nucleon cross section.
Therefore, one can Taylor-expand the integrand in Eq.~(\ref{scf6}) 
around $\sigma_{{\rm tot}}$ and keep only first two 
non-vanishing terms
\begin{equation}
\sigma_{DD}^{hA} \approx \frac{\omega_{\sigma}\,\sigma_{{\rm tot}}^2}{4}
\int d^2 b \,T^2_A(b) \,e^{-\sigma_{{\rm tot}}\, T_A(b)}  \,.
\label{eq:approx}
\end{equation}
In Eq.~(\ref{eq:approx}), $\omega_{\sigma}$ is the energy-dependent parameter, which is proportional to the proton-proton diffractive dissociation cross section 
and which controls  the magnitude of cross section fluctuations
\begin{equation}
\omega_{\sigma}=\frac{\int d \sigma\, \sigma^2\, P(\sigma)}{\left[\int d \sigma\,
\sigma\, P(\sigma) \right]^2}-1 \,.
\label{eq:omega}
\end{equation}

Equation~(\ref{scf6}) can be interpreted as follows. The incoming proton 
diffractively dissociates on the front face of the target nucleus.
The corresponding scattering amplitude squared is proportional to $\omega_{\sigma}\,\sigma_{{\rm tot}}^2\,T^2_A(b)$. 
On the way through the nucleus, the produced diffractive state interacts with all
nucleons of the target and becomes partially absorbed (suppressed).
The corresponding soft suppression factor can be read off from 
Eq.~(\ref{eq:approx}),
\begin{equation}
T_{{\rm soft}}^{\,pA}=\exp\left(-\sigma_{{\rm tot}}\, T_A(b) \right) \,.
\label{eq:T_soft}
\end{equation}
Note that since we have assumed that the dispersion of $P(\sigma)$ around
$\sigma=\sigma_{{\rm tot}}$ is small, the soft suppression factor depends only on 
$\sigma_{{\rm tot}}$.

\subsection{Hard coherent diffraction in DIS on nuclear targets}

Inclusive and coherent DIS on nuclear targets measure usual and diffractive
nuclear parton distribution functions (PDFs), respectively.
The theory of leading twist nuclear shadowing
of usual and diffractive nuclear PDFs
is based on the Gribov's relation between 
nuclear shadowing and diffraction~\cite{Gribov:1968jf,Frankfurt:1998ym},
 Collins' factorization 
theorem for
hard diffraction in DIS~\cite{Collins:1997sr} and the QCD analysis of the 
HERA data on
hard diffraction in DIS on hydrogen~\cite{Breitweg:1998gc,Adloff:1997sc,Adloff:2000qi,unknown:2006hy,:2006hx},
see~\cite{Frankfurt:2003zd} for the review and references.

According to this approach, the nuclear shadowing correction, $\delta f_{j/A}$,
to the nuclear PDF of the flavor $j$, $f_{j/A}=Af_{j/N}-\delta f_{j/A}$,
is expressed in terms of the proton diffractive PDF  $f_{j/N}^{D(4)}$~\cite{Frankfurt:1998ym,Frankfurt:2003gx}
\begin{eqnarray}
\delta x f_{j/A}(x,Q_0^2)&=& 8 \pi \Re e \Bigg[
\frac{(1-i\eta)^2}{1+\eta^2} \int d^2 b \int^{\infty}_{-\infty} dz_1
 \int^{\infty}_{z_1} dz_2 \int_x^{x_0} d x_{\Pomeron} \beta f_{j/N}^{D(4)}(x,Q_0^2,x_{\Pomeron},t=0) \nonumber\\
& \times& \rho_A(b,z_1)\rho_A(b,z_2) \,e^{i x_{\Pomeron} m_N(z_1-z_2)}\,
e^{-\frac{1-i\eta}{2} \sigma_{\rm eff}^j(x,Q_0^2) \int^{z_2}_{z_1}
dz^{\prime} \rho_A(b,z^{\prime})} \Bigg]
\,.
\label{eq:inclusive}
\end{eqnarray}
In Eq.~(\ref{eq:inclusive}),
$x$ and $Q_0^2$ are the Bjorken variables;
$x_{\Pomeron}$ is the target longitudinal momentum fraction loss
or the longitudinal
momentum fraction carried by the diffractive exchange (the "Pomeron");
$\beta=x/x_{\Pomeron}$; $x_0=0.1$;
 $\sigma_{{\rm eff}}^j$ is the effective
rescattering cross section of the intermediate diffractive state;
$\eta$ is the ratio of the real to the imaginary parts of the 
elementary diffractive amplitude. In the present analysis, 
$\eta=\pi/2 \,(\alpha_{\Pomeron}(0)-1)=0.185$~\cite{unknown:2006hy,:2006hx}.

It is important to mention that since the $t$-dependence of the nuclear 
form factor is much steeper than that of the nucleon diffractive structure function, it is a good approximation to use $f_{j/N}^{D(4)}(x,Q_0^2,x_{\Pomeron},t_{{\rm min}})$ 
[$t_{{\rm min}} \approx -(x_{\Pomeron} m_N)^2$]
instead of 
$f_{j/N}^{D(4)}(x,Q_0^2,x_{\Pomeron},t)$ in Eq.~(\ref{eq:inclusive}).
Moreover, in the considered range of Bjorken $x$, $t_{{\rm min}}$ is small enough
so that $f_{j/N}^{D(4)}(x,Q_0^2,x_{\Pomeron},t_{{\rm min}})$ and $f_{j/N}^{D(4)}(x,Q_0^2,x_{\Pomeron},t=0)$ practically coinside.

As follows from Eq.~(\ref{eq:inclusive}), $\sigma_{{\rm eff}}^j$  determines the nuclear correction to $\delta f_{j/A}$ due to the interaction with
 two and more nucleons (the interaction associated with the
rescattering of the intermediate diffractive state). 
This cross section is defined in terms of the nucleon
diffractive ($f_{j/N}^{D(4)}$) and usual ($f_{j/N}$) PDFs~\cite{Frankfurt:1998ym,Frankfurt:2003zd}
\begin{equation}
\sigma_{{\rm eff}}^j(x,Q_0^2)=\frac{16 \pi}{xf_{j/N}(x,Q_0^2)}
\int_x^{x_0} d x_{\Pomeron} \,\beta f_{j/N}^{D(4)}(\beta,Q_0^2,x_{\Pomeron},t=0) \,.
\label{eq:sigmaeff}
\end{equation}
\begin{figure}[t]
\begin{center}
\epsfig{file=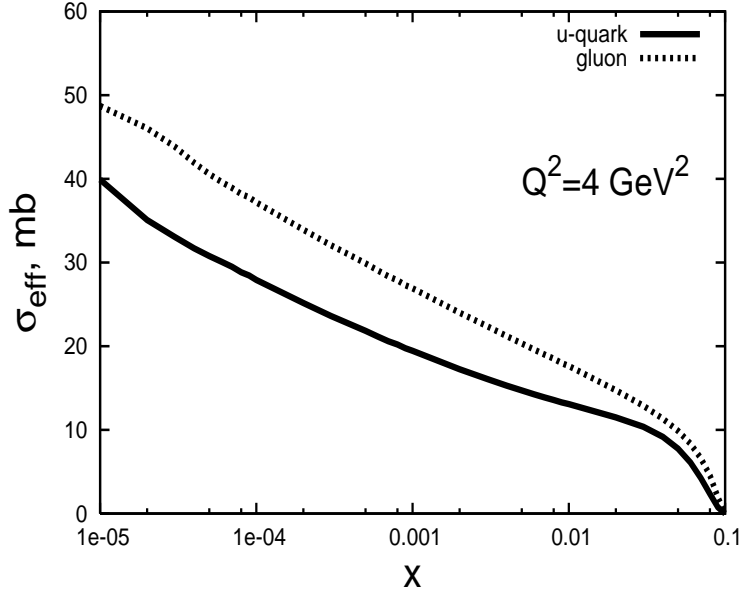,width=10cm,height=8cm}
\caption{The effective cross section $\sigma_{{\rm eff}}^j$ of Eq.~(\ref{eq:sigmaeff}) for ${\bar u}$-quarks and gluons 
at $Q_0^2=4$ GeV$^2$.}
\label{fig:sigma_h12006_2jets}
\end{center}
\end{figure}
Figure~\ref{fig:sigma_h12006_2jets} presents $\sigma_{{\rm eff}}^j$ for
gluons and ${\bar u}$-quarks as a function of $x$ at $Q_0^2=4$ GeV$^2$.
We used the recent QCD analysis of H1
 data on hard diffraction at HERA~\cite{unknown:2006hy,:2006hx} and
CTEQ5M fit to inclusive PDFs~\cite{Lai:1999wy}.

Also, since Eq.~(\ref{eq:sigmaeff}) involves the nucleon diffractive PDFs
at $t = 0$, one has to make an assumption about the
$t$-dependence of the nucleon diffractive PDFs. 
Experimentally, the $t$-dependence of the diffractive cross section is found to be practically constant  as a function of $\beta$, while the contribution of the gluon diffractive PDF increases strongly with a decrease of $\beta$~\cite{unknown:2006hy,:2006hx}.
Hence, 
in this work, we assume that
all PDFs have the same exponential $t$-dependence,
\begin{equation}
f_{j/N}^{D(4)}(\beta,Q_0^2,x_{\Pomeron},t)=e^{-B_{\rm diff}ŧ|t|} f_{j/N}^{D(4)}(\beta,Q_0^2,x_{\Pomeron},t=0) \,,
\label{eq:diffractive_slope}
\end{equation}
where $B_{\rm diff}=6$ GeV$^{-2}$ is taken from the recent H1 measurement 
with the leading proton spectrometer~\cite{:2006hx}.

\begin{figure}[t]
\begin{center}
\epsfig{file=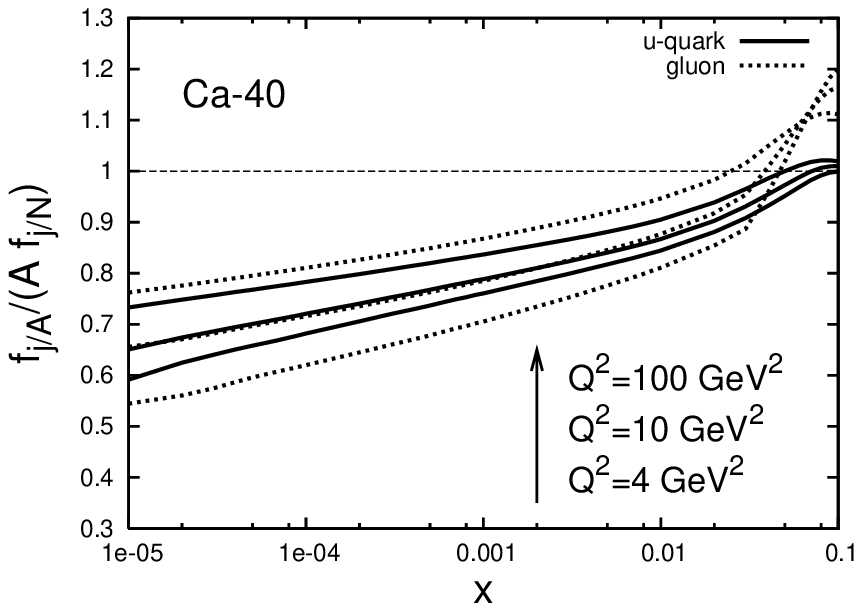,width=8cm,height=8cm}
\epsfig{file=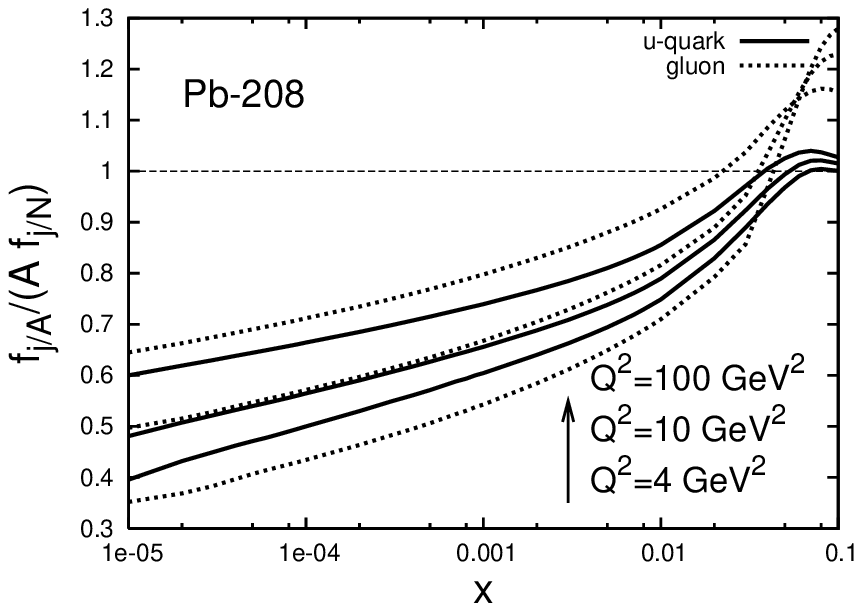,width=8cm,height=8cm}
\caption{The ratio of the nuclear to nucleon PDFs, $f_{j/A}/(A  f_{j/N})$,
for $^{40}$Ca  and $^{208}$Pb at $Q^2=4$, 10 and 100 GeV$^2$ as a 
function of Bjorken $x$, see Eq.~(\ref{eq:inclusive}).
 The solid curve corresponds to ${\bar u}$-quarks; the dotted curve
corresponds to gluons.}
\label{fig:ca40_input_2jets}
\end{center}
\end{figure}

Figure~\ref{fig:ca40_input_2jets} presents an example of our calculation of
nuclear shadowing for nuclear PDFs of $^{40}$Ca and $^{208}$Pb
as a function of $Q^2$ and Bjorken $x$. The solid curve corresponds to
the ratio $f_{j/A}/(A  f_{j/N})$ for ${\bar u}$-quarks;
the dotted curve corresponds to gluons.
The lower set of curves corresponds to $Q^2=Q_0^2=4$ GeV$^2$. In addition to
nuclear shadowing given by Eq.~(\ref{eq:inclusive}), we have introduced an 
enhancement (antishadowing) of nuclear gluon PDF on the interval 
$0.03 \leq x \leq 0.2$, which is modelled by requiring the conservation of the
momentum sum rule, see e.g.~\cite{Frankfurt:2003zd}. 
The two other sets of predictions for  $Q^2=10$ GeV$^2$ and $Q^2=100$ GeV$^2$
are obtained by NLO DGLAP evolution.

We would like to point out that the numerical analysis of nuclear shadowing
presented in this work differs from our earlier analyses, see e.g.~\cite{Frankfurt:2003zd}, 
because we now use the most recent H1 fits to
nuclear diffractive PDFs and a different value of $B_{\rm diff}$, see
Eq.~(\ref{eq:diffractive_slope}). However, the changes in the predicted 
nuclear shadowing are not large. The ratio $f_{j/A}/(A  f_{j/N})$ for
${\bar u}$-quarks in Fig.~\ref{fig:ca40_input_2jets} is very similar to
our earlier result~\cite{Frankfurt:2003zd}.
In the gluon channel,
the ratio $f_{j/A}/(A  f_{j/N})$ 
in Fig.~\ref{fig:ca40_input_2jets} is similar to the 
lower-gluon-shadowing scenario of~\cite{Frankfurt:2003zd}.

Next we turn to nuclear diffractive PDF.
In the Glauber-Gribov approach,
the nuclear diffractive PDF of the flavor $j$, $f_{j/A}^{D(3)}$,
is expressed in terms of the proton diffractive PDF  $f_{j/N}^{D(4)}$ as follows~\cite{Frankfurt:2003gx} 
\begin{eqnarray}
x f_{j/A}^{D(3)}(x,Q_0^2,x_{\Pomeron})&=&4\,\pi \beta\,f_{j/N}^{D(4)}(x,Q_0^2,x_{\Pomeron},t=0) \int d^2 b \nonumber\\
&&\times \left| \int^{\infty}_{-\infty} dz \,e^{i x_{\Pomeron} m_N z} e^{-\sigma_{{\rm eff}}^j(x,Q_0^2)/2 \int_{z}^{\infty} d z^{\prime}\rho_A(b,z^{\prime})} \rho_A(b,z)\right|^2 
\nonumber\\
& \approx & 16\,\pi f_{j/N}^{D(4)}(x,Q_0^2,x_{\Pomeron},t=0) \int d^2 b 
\left(\frac{1-e^{-\sigma_{{\rm eff}}^j(x,Q_0^2)/2 \,T_A(b)}}{\sigma_{{\rm eff}}^j(x,Q_0^2)}\right)^2
\,.
\label{eq:masterD}
\end{eqnarray}
The last line is an approximation valid at small $x_{\Pomeron}$, when the
effect of the coherence length [the factor $\exp(i x_{\Pomeron} m_N z)$] can be 
neglected.
In the opposite limit of large $x_{\Pomeron}$, $x_{\Pomeron} \geq 0.05$, 
the dominant contribution to the nuclear diffractive structure function is given by
the impulse approximation, i.e.~by Eq.~(\ref{eq:masterD}) where
$\sigma_{{\rm eff}}^j$ is set to zero.
In Eq.~(\ref{eq:masterD}),
the superscripts $(3)$ and $(4)$ denote the dependence on three and four 
variables, respectively.
Note that similarly to Eq.~(\ref{eq:approx}), we neglected the  slope and
 the real part
of the elementary diffractive amplitude in Eq.~(\ref{eq:masterD}).

One can quantify nuclear diffractive PDFs by introducing the
probability of diffraction for a given parton flavor $j$, $P^j_{{\rm diff}}$~\cite{Frankfurt:1998ym},
\begin{equation}
P^j_{{\rm diff}}=\frac{\int_x^{x_0} dx_{\Pomeron}\, x f_{j}^{D(3)}(x,Q_0^2,x_{\Pomeron})}{x f_{j}(x,Q_0^2)} \,.
\label{eq:probability}
\end{equation}

\begin{figure}[t]
\begin{center}
\epsfig{file=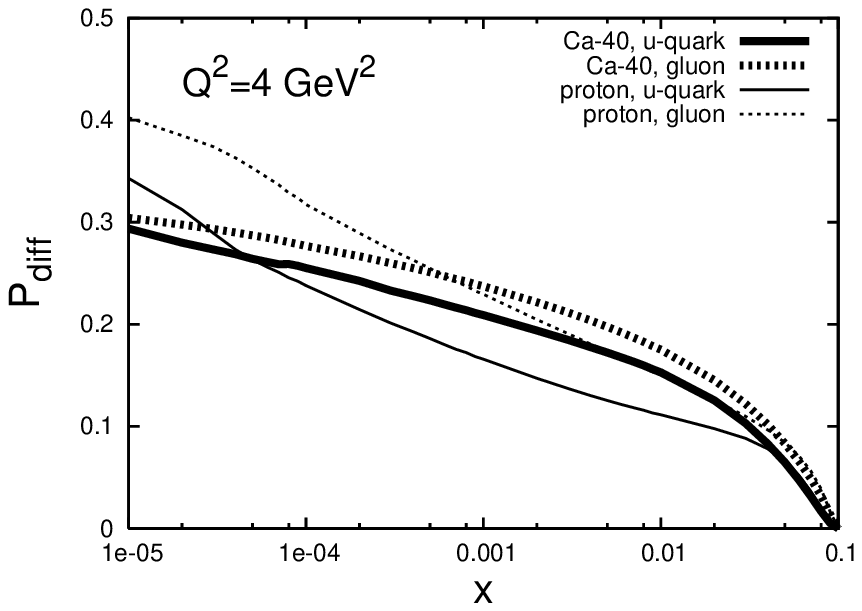,width=8cm,height=8cm}
\epsfig{file=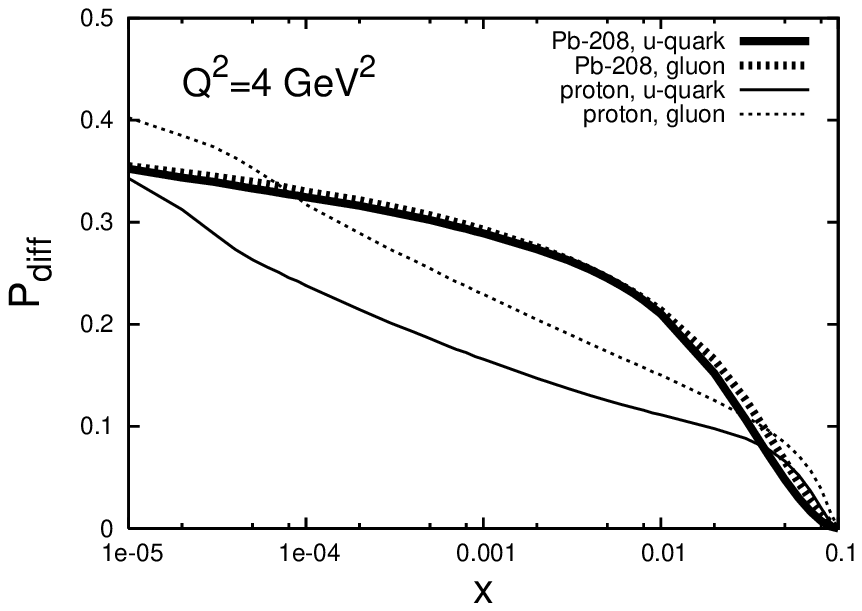,width=8cm,height=8cm}
\caption{The probability of hard diffraction in DIS, $P^j_{{\rm diff}}$,
see Eq.~(\ref{eq:probability}). The left panel corresponds to DIS on $^{40}$Ca;
the right panel corresponds to DIS on $^{208}$Pb.
For comparison, the results for DIS on hydrogen are given by thin curves.
 The solid curves correspond to the ${\bar u}$-quark channel; the dotted curves
correspond to the gluon channel.
}
\label{fig:probability}
\end{center}
\end{figure}

An example of the evaluation of the probability of hard diffraction in DIS 
according to Eq.~(\ref{eq:probability}) is 
presented in Fig.~\ref{fig:probability}, where  $P^j_{{\rm diff}}$ is plotted
at fixed $Q_0^2=4$ GeV$^2$ as a function of Bjorken $x$.
The left panel corresponds to DIS on $^{40}$Ca;
the right panel corresponds to DIS on $^{208}$Pb.
For comparison, the results for DIS on hydrogen are also given by thin curves.
The solid curves correspond to the ${\bar u}$-quark channel; the dotted curves
correspond to the gluon channel.

We would like to point out the following two features of $P^j_{{\rm diff}}$
presented in Fig.~\ref{fig:probability}. First, the difference between the quark and
the gluon channels is very small. While the quark and gluon diffractive and 
usual nuclear PDFs are different, their difference cancels to a large extent
in the ratio $P^j_{{\rm diff}}$ (the cancellation is larger for heavier nuclei). 
Second, even for such a heavy nucleus as $^{208}$Pb, $P^j_{{\rm diff}} \leq 0.36$, which should be compared to the asymptotic ($A \to \infty$ and $\sigma_{{\rm eff}}^j \to \infty$) upper limit  $P^j_{{\rm diff}} =0.5$. An examination shows that while
close to the center of the nucleus ($b \approx 0$), 
the probability of diffraction is very
close to $1/2$, the 
contribution of the nuclear edge significantly dilutes $P^j_{{\rm diff}}$.

Equation~(\ref{eq:masterD})  can be interpreted as follows. 
The incoming virtual photon fluctuates into its hard diffractive 
component long time before the photon interacts with the target
(we ignore the effect of the 
finite coherent length). The hard diffractive component elastically rescatters on 
the target nucleus, which gives the suppression factor $[1-\exp(-\frac{1}{2}\sigma_{{\rm eff}}^j(x,Q_0^2)\, T_A(b)]^2$, and emerges as the
final hard diffractive state. 

One should note that the approximate expression for 
$f_{j/A}^{D(3)}$ [the last line in Eq.~(\ref{eq:masterD})] corresponds to the first term
in Eq.~(\ref{eq:approx}) since the elastic contribution to DIS is absent (suppressed by the 
smallness of $\alpha_{{\rm e.m.}}$). 
Therefore, the analogy between Eqs.~(\ref{eq:approx}) and (\ref{eq:masterD})
enables us to introduce the attenuation factor characterizing the suppression of
hard coherent diffraction in DIS on nuclear targets
due to nuclear shadowing,
\begin{equation}
T_{{\rm hard}}^{\,\gamma^{\ast} A}=\exp\left(-\sigma_{{\rm eff}}^j(x,Q_0^2)\, T_A(b) \right) \,.
\label{eq:T_hard}
\end{equation}

\begin{figure}[t]
\begin{center}
\epsfig{file=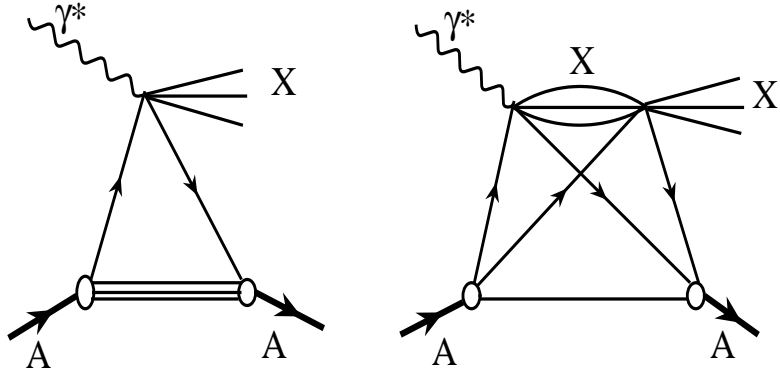,width=12cm,height=7cm}
\caption{Feynman graphs representing the first two terms of the multiple
scattering series~(\ref{eq:masterD}) for the nuclear diffractive
parton distribution $f_{j/A}^{D(3)}$.}
\label{fig:2jets_inclusive_diffraction_FG}
\end{center}
\end{figure}

The graphical representation of Eq.~(\ref{eq:masterD}), when only
the interaction with one and two nucleons of the target is retained, is shown in Fig.~\ref{fig:2jets_inclusive_diffraction_FG}.
The right graph helps to understand why  
$T_{{\rm hard}}^{\,\gamma^{\ast} A}$ is driven by 
$\sigma_{{\rm eff}}^j$.
As seen from the graph, 
the strength of the hard rescattering is defined by the 
$XN \to XN$ cross section summed over all $X$. This cross section
is nothing but the ratio of the $\gamma^{\ast} N \to XN$ to the $\gamma^{\ast} N \to X$
cross sections summed over $X$, which corresponds exactly to
 $\sigma_{{\rm eff}}^j$ defined by Eq.~(\ref{eq:sigmaeff}).

It is important to emphasize that, in general, the 
calculation of $T_{{\rm hard}}^{\,\gamma^{\ast} A}$
is model-independent only for the interaction
with one or two nucleons. For the interaction with $N \ge 3 $ nucleons,
 we implicitly used the so-called quasi-eikonal approximation
in Eq.~(\ref{eq:T_hard}), 
which assumes that the diffractively
produced state elastically rescatters on the nucleons. This
approximation is equivalent to the observation of the small dispersion of $P(\sigma)$
used in the derivation of Eq.~(\ref{eq:approx}).

\subsection{Hard coherent proton-nucleus diffraction}

As an example of hard coherent diffractive processes on heavy 
nuclear targets, we consider
the hard coherent diffractive production of two jets in the reaction
$p+A \to 2\,{\rm jets}+X+A$. In this process, $A$ denotes the nucleus; 
$X$ denotes the soft diffractive component; 
the invariant mass of the jets provides the hard scale.

The cross section of the $p+A \to 2\,{\rm jets}+X+A$ reaction
can be readily obtained by 
generalizing the well-known expression for
the dijet inclusive cross section in hadron-hadron scattering~\cite{Ellis} 
and by introducing the new quantity, the screened nuclear diffractive PDFs
$\tilde{f}_{j/A}^{D(3)}$,
\begin{eqnarray}
&&\frac{d^3 \sigma^{p+A \to 2\,{\rm jets}+X+A}}{d x_1 \,d p_T^2 \,d x_{\Pomeron}} 
\nonumber\\
&& \propto \sum_{\substack{i,j,\\k,l=q,\bar{q},g}} f_{i/p}(x_1,Q_{{\rm eff}}^2)\tilde{f}_{j/A}^{D(3)}(x_2,Q_{{\rm eff}}^2,x_{\Pomeron}) \overline{\sum}|{\cal M}(ij \to kl)|^2 \frac{1}{1+\delta_{kl}} 
\,,
\label{eq:cs_hard}
\end{eqnarray}
where $f_{i/p}$ are the usual proton PDFs;
 $\overline{\sum}|{\cal M}(ij \to kl)|^2$ are the invariant matrix elements for
two-to-two parton scattering given in Table~7.1 of~\cite{Ellis};
$x_1$ and $x_2$ are the light-cone momentum fractions of the proton
and the nucleus active quarks, respectively;
$p_T$ is the transverse momentum of each of the final jets;
$Q_{{\rm eff}}$ is the effective hard scale of the process.
For the simplification of our analysis, 
we consider only the case of $90^{0}$ hard
 parton scattering in the center of mass, which constrains $x_1$ (as a function
of $x_2=\beta \,x_{\Pomeron}$) and 
$Q_{{\rm eff}}^2$
\begin{equation} 
x_1=\frac{4 \, p_T^2}{\beta x_{\Pomeron} \,s} \,, \quad \quad Q_{{\rm eff}}^2=4\, p_T^2 \,, 
\end{equation} 
where $\sqrt{s}$ is the proton-nucleon invariant energy.
The term ''screened PDF'' means that this parton distribution contains
certain soft suppression effects, i.e.~the screened PDF is suppressed compared to
the analogous PDF extracted from hard processes.

\begin{figure}[t]
\begin{center}
\epsfig{file=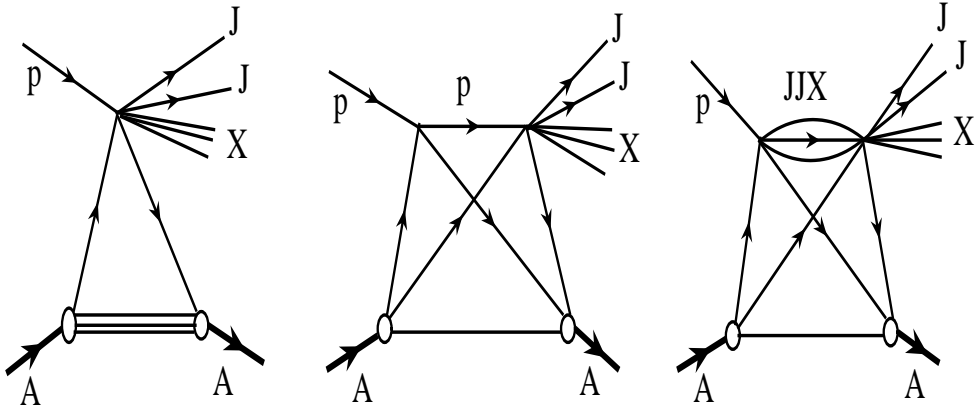,width=14cm,height=9cm}
\vskip -2cm
\caption{Feynman graphs representing the first two terms of the multiple
scattering series for the $p+A \to 2\,{\rm jets}+X+A$ scattering amplitude.}
\label{fig:2jets_FG}
\end{center}
\end{figure}

The derivation of the expression for the screened nuclear diffractive PDFs, $\tilde{f}_{j/A}^{D(3)}$, is carried out
similarly to the derivation of Eq.~(\ref{eq:approx})
 [see also Fig.~\ref{fig:2jets_FG}]
\begin{equation}
\tilde{f}_{j/A}^{D(3)}(x,Q_0^2,x_{\Pomeron})
 \approx  4\,\pi \tilde{f}_{j/N}^{D(4)}(x,Q_0^2,x_{\Pomeron},t=0) \int d^2 b \,T_A^2(b)\,
e^{-(\sigma_{{\rm tot}}(s)+\sigma_{{\rm eff}}^j(x,Q_0^2)) T_A(b)}
\,,
\label{eq:npdf_effective}
\end{equation} 
where $\tilde{f}_{j/N}^{D(4)}$ is the screened diffractive PDF of the nucleon, 
which enters the QCD description of the $p+p \to 2\,{\rm jets}+X+p$ reaction;
$\sigma_{{\rm tot}}$ is the total proton-nucleon cross section;
$\sigma_{{\rm eff}}^j$ is the effective rescattering cross section of 
Eq.~(\ref{eq:sigmaeff}). In Eq.~(\ref{eq:npdf_effective}), we neglected the
slope and the real part of the elementary 
$p+N \to 2\,{\rm jets}+X+N$
scattering amplitude and a small longitudinal momentum
transfer in the  $p+N \to 2\,{\rm jets}+X+N$ vertex.

It is important to emphasize that in the case of hard coherent proton-nucleus
diffraction, the nuclear suppression factor, $T_{{\rm hard}}^{\,pA}$, is a 
product of the soft and hard suppression factors introduced previously,
\begin{equation}
T_{{\rm hard}}^{\,pA}=T_{{\rm soft}}^{\,pA}\ T_{{\rm hard}}^{\,\gamma^{\ast}A} \,.
\label{eq:T_hard_pa}
\end{equation}
This can be understood from Fig.~\ref{fig:2jets_FG}, which represents
the single and double scattering contributions to the 
$p+A \to 2\,{\rm jets}+X+A$ scattering amplitude. 
The rescattering cross section of the middle graph is 
$\sigma_{{\rm tot}}$; the rescattering cross section of the right
graph is $\sigma_{{\rm eff}}^j$ (we assume that all diffractive intermediate
states correspond to the same rescattering cross section).
Therefore, the resulting nuclear 
attenuation, which results from the sum of the middle and right graphs, is 
driven by the $\sigma_{{\rm tot}}+\sigma_{{\rm eff}}^j$ cross section.

Equation~(\ref{eq:npdf_effective}) can be interpreted 
in two complimentary ways.
On the one hand, one can start from soft diffractive dissociation
of protons on heavy nuclei, see Eq.~(\ref{eq:approx}).
Since we are interested in the hard diffractive component of the 
diffractive dissociation cross section, one has to take into account
the additional suppression of nuclear diffractive PDFs given by 
$T_{{\rm hard}}^{\,\gamma^{\ast}A}$. As a result, one arrives at 
Eq.~(\ref{eq:T_hard_pa}).
On the other hand, one can 
start from the expression for  inclusive diffraction of protons
on nuclei, which is proportional to the nuclear diffractive PDFs~(\ref{eq:masterD}).
Since the final diffractive state contains a soft component, which is 
partially absorbed by the nucleus, one should take into account this suppression by
introducing  the factor $T_{{\rm soft}}^{\,pA}$, which represents the probability 
of the absence of soft inelastic interactions at a given impact parameter $b$.

We quantify the suppression of the nuclear screened diffractive PDFs $\tilde{f}_{j/A}^{D(3)}$
compared to the nucleon screened diffractive PDFs $\tilde{f}_{j/N}^{D(3)}$
by introducing the factor $\lambda^j$
\begin{eqnarray}
\lambda^j(x,Q^2)& \equiv & \frac{\tilde{f}_{j/A}^{D(3)}(\beta,Q^2,x_{\Pomeron})}{\tilde{f}_{j/N}^{D(3)}(\beta,Q^2,x_{\Pomeron})}=4\,\pi \frac{\tilde{f}_{j/N}^{D(4)}(x,Q_0^2,x_{\Pomeron},t=0)}{\tilde{f}_{j/N}^{D(3)}(\beta,Q^2,x_{\Pomeron})} \int d^2 b \,T_A^2(b)\,
e^{-(\sigma_{{\rm tot}}(s)+\sigma_{{\rm eff}}^j(x,Q^2)) T_A(b)} \nonumber\\
&=&
B_{{\rm diff}}(4 \pi) \int d^2 b \,T_A^2(b)\, \exp\left[-(\sigma_{{\rm tot}}(s)+\sigma_{{\rm eff}}^j(x,Q^2)) T_A(b)\right] \,,
\label{eq:lambda_parton}
\end{eqnarray}
where in the last line we introduced the slope of the $t$-dependence of the screened
diffractive PDFs,
\begin{equation}
\tilde{f}_{j/N}^{D(4)}(\beta,Q^2,x_{\Pomeron},t)=\exp(-B_{{\rm diff}} \, |t|) \,
\tilde{f}_{j/N}^{D(4)}(\beta,Q^2,x_{\Pomeron},t=0) \,.
\label{eq:diffractive_slope2}
\end{equation}
Note also that $\tilde{f}_{j/N}^{D(3)}(\beta,Q^2,x_{\Pomeron}) \equiv \int dt \tilde{f}_{j/N}^{D(4)}(\beta,Q^2,x_{\Pomeron},t)$.

\begin{figure}[t]
\begin{center}
\epsfig{file=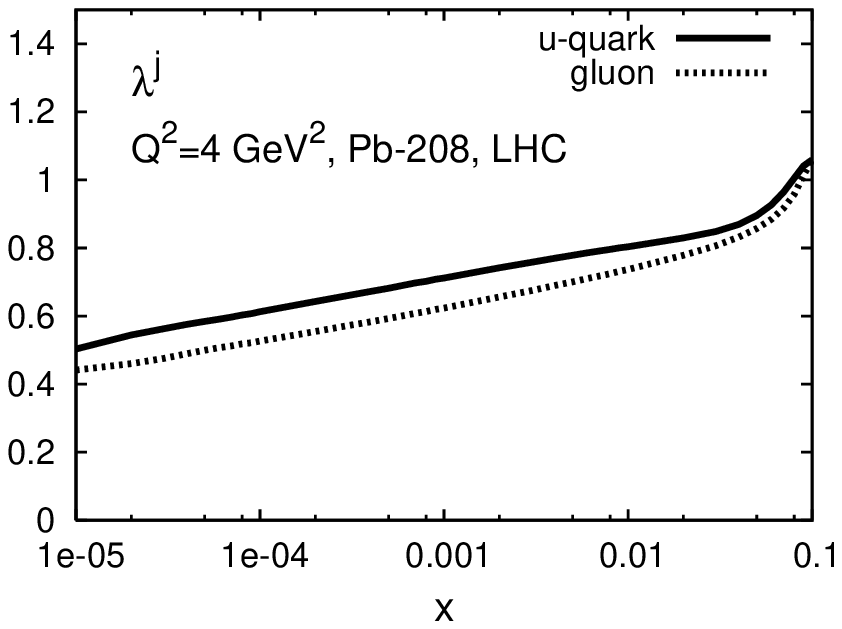,width=8cm,height=8cm} 
\epsfig{file=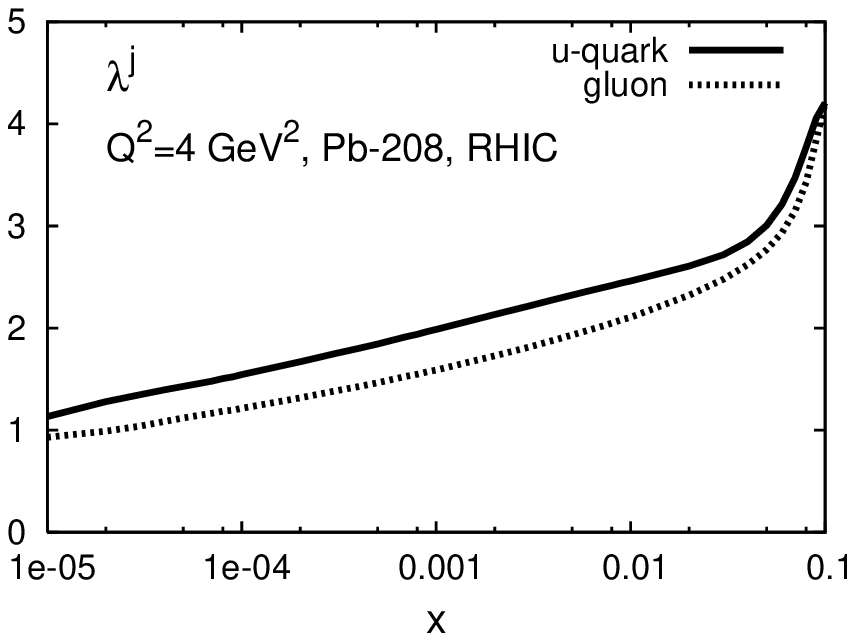,width=8cm,height=8cm} 
\caption{The suppression factor $\lambda^j$ of Eq.~(\ref{eq:lambda_parton}) as
a function of Bjorken $x$ at $Q^2=4$ GeV$^2$
in the LHC ($\sqrt{s}=8.8$ TeV) and RHIC ($\sqrt{s}=200$ GeV) kinematics.
 The solid curves correspond to the $\bar{u}$-quark;
the dotted curves correspond to the gluons.}
\label{fig:lambda_parton}
\end{center}
\end{figure}

Certain features of Eq.~(\ref{eq:lambda_parton}) deserve a discussion.
First,
while the diffractive PDFs depend separately on $\beta$ and
  $x_{\Pomeron}$, the
suppression factor $\lambda^j$ depends only on their
 product $x=\beta \,x_{\Pomeron}$
in our approach. Second, at
 the LHC energies, where $\sigma_{{\rm tot}} $ is of the order of
 100 mb,
the dependence of $\lambda^j$ on $\sigma_{{\rm eff}}^j$ is rather weak.
Therefore, we expect that $\lambda^j$ is rather similar for 
different parton flavors $j$.
In addition, since the slope $B_{{\rm diff}}$ is independent on the 
hard scale $Q^2$ in our approach, $\lambda^j$ has very weak dependence on $Q^2$, 
which enters only through the $Q^2$-dependence of $\sigma_{{\rm eff}}^j$.

In our numerical analysis of Eq.~(\ref{eq:lambda_parton}), we used the following input.
The slope of the $t$-dependence of $\tilde{f}_{j/N}^{D(4)}$
was taken from the recent H1 measurement of hard inclusive diffraction in
DIS on hydrogen, 
$B_{{\rm diff}}=6$ GeV$^{-2}$~\cite{:2006hx}.

The total proton-nucleon scattering cross section, $\sigma_{{\rm tot}}$,
was taken from~\cite{Donnachie:1992ny}
\begin{equation}
\sigma_{{\rm tot}}(s)=21.7\,s^{0.0808}+56.08\,s^{-0.4525} \,.
\label{eq:donnachie}
\end{equation}

The effective cross section $\sigma_{{\rm eff}}^j$ was evaluated using
Eq.~(\ref{eq:sigmaeff}) and the recent QCD fits to the H1 measurement
of hard inclusive diffraction on hydrogen~\cite{unknown:2006hy,:2006hx},
see Fig.~\ref{fig:sigma_h12006_2jets}.

For the nucleon density $\rho_A$, the two-parameter Fermi 
model was used~\cite{DeJager:1987qc}.

\begin{figure}[t]
\begin{center}
\epsfig{file=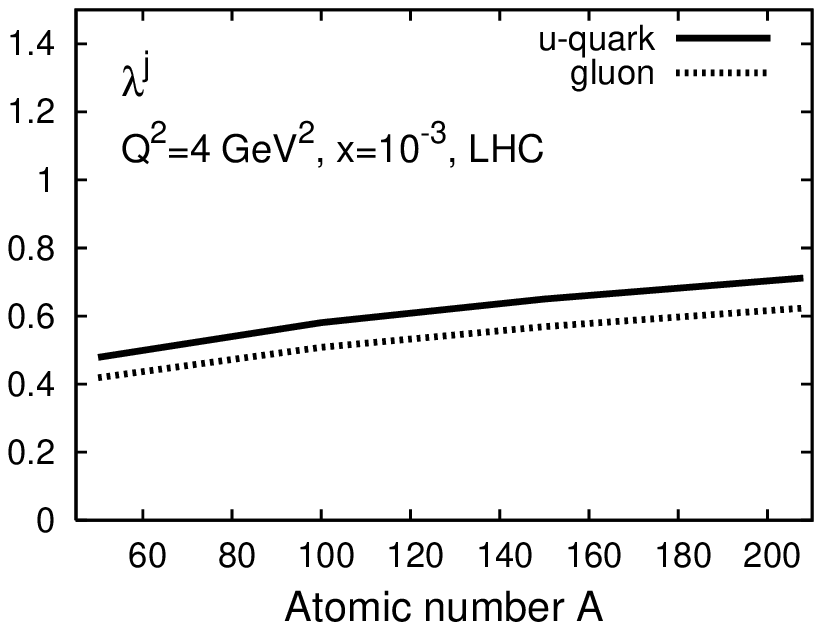,width=8cm,height=8cm}
\epsfig{file=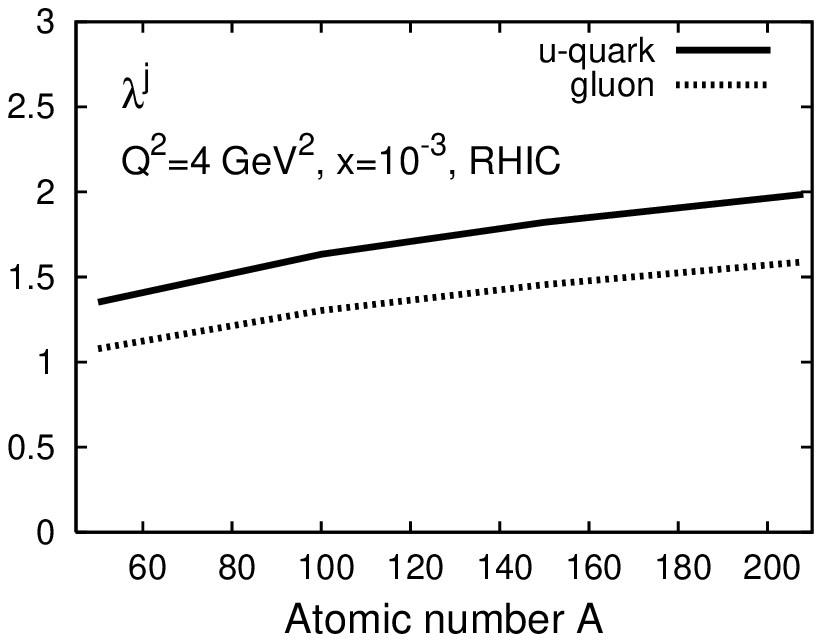,width=8cm,height=8cm}
\caption{The suppression factor $\lambda^j$ of Eq.~(\ref{eq:lambda_parton}) as
a function of the atomic number at $x=10^{-3}$ and at $Q^2=4$ GeV$^2$
in the LHC ($\sqrt{s} \approx 9$ TeV) and RHIC ($\sqrt{s}=200$ GeV) kinematics. 
The labeling of the curves is the same as in
Fig.~\ref{fig:lambda_parton}.
}
\label{fig:lambda_parton_adep}
\end{center}
\end{figure}

Figures~\ref{fig:lambda_parton} and \ref{fig:lambda_parton_adep} present
the results of our calculations of $\lambda^j$ in the 
LHC and RHIC kinematics.
The LHC kinematics corresponds to $\sqrt{s} \approx 9$ TeV per nucleon for proton-nucleus
collisions~\cite{Morsch}; the RHIC kinematics corresponds to $\sqrt{s}=200$ GeV.
Figure~\ref{fig:lambda_parton} 
gives $\lambda^j$ as a function of
 Bjorken $x$ 
at $Q^2=4$ GeV$^2$.
 The solid curves correspond to the $\bar{u}$-quark;
the dotted curves correspond to the gluons.
Despite the fact that  $\lambda^j$ is of the order of unity at the
LHC and of the order of several units at the RHIC energies, the 
corresponding suppression of hard diffraction is very large because in the
absence of the suppression, nuclear diffractive PDFs are enhanced compared to
the nucleon diffractive PDFs by the factor
 $f_{j/A}^{D(3)}/f_{j/N}^{D(3)} \propto A^{4/3}$.

Figure~\ref{fig:lambda_parton_adep} presents the $A$-dependence of  
$\lambda^j$ at $x=10^{-3}$ and at $Q^2=4$ GeV$^2$, i.e.~at fixed $\sigma_{{\rm eff}}^j$.
 As seen from
Fig.~\ref{fig:lambda_parton_adep}, the 
$A$-dependence of $\lambda^j$ is rather slow. A simple fit gives
that $\lambda^j \propto A^{0.28}$ at the LHC and RHIC.

\section{Hard diffraction and ultraperipheral proton-nucleus collisions}
\label{sec:ultraperipheral}

\begin{figure}[t]
\begin{center}
\epsfig{file=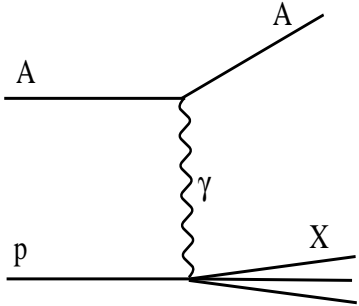,width=8cm,height=8cm}
\caption{The ultraperipheral $p+A \to X+A$ scattering.}
\label{fig:ultraperipheral}
\end{center}
\end{figure}

In proton-heavy nucleus ($^{208}$Pb, for example) collisions,
most of the diffractive events ($\sim 80\%$) will be generated by the
scattering of the proton off the coherent nuclear Coulomb field at large
 impact parameters,
$p+A \to p+\gamma+A \to X+A$~\cite{Guzey:2005tk}, see
Fig.~\ref{fig:ultraperipheral}.
These ultraperipheral proton-nucleus
collisions open a possibility for studies of hard photon-proton interactions
 at extremely high energies 
and allow one to probe the gluon density in the proton at the values of 
Bjorken $x$, which are a factor of ten smaller (for the same virtuality)
 than those probed at HERA~\cite{Baur:2001jj,Bertulani:2005ru,Strikman:2005yv}.

In this Section, we estimate the ratio of  the jet production 
in hard coherent proton-heavy nucleus diffraction 
($p+A \to 2\,{\rm jets}+X+A$) to  the production
of hard jets by the photon-proton interaction, where the photon is 
coherently produced by the elastically recoiled nucleus,
$p+A  \to p+\gamma+A \to 2\,{\rm jets}+X+A$.
This corresponds to the situation when the generic final state $X$ in
Fig.~\ref{fig:ultraperipheral} contains a hard two-jet component and
a soft remaining part $X$.

Qualitatively, we expect that the ratio of the hard dijet
production due to these two mechanisms
will be rather small because of the 
following two suppression effects.

First, hard diffractive dijet production is suppressed by the 
factor discussed in Sect.~\ref{sec:formula}.
 Second, the shapes of the parton distribution in the photon and 
in the screened nuclear 
diffractive PDFs are rather different. In the photon case, 
the  dominant contribution to the photon PDFs comes from $\beta \sim 1$
 corresponding to the kinematics where a pair of jets is at the rapidities
 close to the gap.
On the other hand, in the screened nuclear diffractive PDFs at large
 virtualities,
which are relevant for the measurements at the LHC,
the main contribution comes from small $\beta$, see Fig.~3 
of~\cite{Frankfurt:2003gx}.

Using the definition of the suppression factor $\lambda^j$~(\ref{eq:lambda_parton}), 
the hard coherent proton-nucleus diffractive dijet
 cross section can be written as [see Eq.~(\ref{eq:cs_hard})] 
\begin{eqnarray}
&&\frac{d^3 \sigma^{p+A \to 2\,{\rm jets}+X+A}}{d x_1 \,d p_T^2 \,d x_{\Pomeron}} 
\nonumber\\
&\propto& r_{{\rm h}} \sum_{\substack{i,j,\\k,l=q,\bar{q},g}} f_{i/p}(x_1,Q_{{\rm eff}}^2)
\lambda^j(\beta x_{\Pomeron},Q_{{\rm eff}}^2)f_{j/N}^{D(3)}(\beta,Q_{{\rm eff}}^2,x_{\Pomeron})
 \overline{\sum}|{\cal M}(ij \to kl)|^2 \frac{1}{1+\delta_{kl}}
\,,
\label{eq:cs_hard2}
\end{eqnarray}
In the second line of Eq.~(\ref{eq:cs_hard2}), we introduced an additional
suppression factor $r_{{\rm h}}$,
\begin{equation}
\tilde{f}_{j/N}^{D(3)}= r_{{\rm h}}\,f_{j/N}^{D(3)} \,,
\label{eq:screen}
\end{equation}
which, according to the discussion in Sect.~\ref{sec:intro},
takes into 
account the significant factorization breaking in hard hadron-hadron diffraction.

In our numerical analysis, we used the following model for the suppression factor
$r_{{\rm h}}$
\begin{equation}
r_{{\rm h}}=\frac{0.75}{N(s)}=0.75\,\left(\int^{0.1}_{1.5/s} dx_{\Pomeron} \int^{0}_{-\infty} dt \, f_{\Pomeron / p}(x_{\Pomeron},t)\right)^{-1} \,.
\label{eq:goulianos}
\end{equation}
This expression is based on the phenomenological model of~\cite{Goulianos:1995wy},
 which describes the suppression of 
 diffraction at the Tevatron ($\sqrt{s}=546$ and 1800 GeV) by 
rescaling the Pomeron flux, $f_{\Pomeron / p}(x_{\Pomeron},t)$, 
by the factor $N(s)$. In Eq.~(\ref{eq:goulianos}), the Pomeron flux is given
by the following expression
\begin{equation}
f_{\Pomeron / p}(x_{\Pomeron},t)=\frac{1}{x_{\Pomeron}^{1+2\,\epsilon+2\,\alpha^{\prime}t}} 
\frac{\beta^2_{\Pomeron pp}(t)}{16 \pi} \,,
\label{eq:pomeron_flux}
\end{equation}
where $\epsilon=0.1$; $\alpha^{\prime}=0.25$ GeV$^{-2}$;
$\beta_{\Pomeron pp}(t)$ is the $\Pomeron pp$ form factor~\cite{Goulianos:1995wy}.

We also introduced the additional factor $0.75$ in Eq.~(\ref{eq:goulianos})
in order to phenomenologically take into account the observation that the effects of
factorization breaking should be larger in the 
elementary diffractive PDFs at $t=0$ [see Eq.~(\ref{eq:lambda_parton})] 
than in the $t$-integrated diffractive PDFs [see Eq.~(\ref{eq:goulianos})].

The application of Eq.~(\ref{eq:goulianos}) at the RHIC and LHC energies gives
\begin{eqnarray}
r_{{\rm h}} &=& \frac{1}{4.2} \,, \quad {\rm RHIC} \,, \nonumber\\
r_{{\rm h}} &=& \frac{1}{16.0} \,, \quad {\rm LHC} \,.
\label{eq:goulianos3}
\end{eqnarray}
Note that as follows from the definition~(\ref {eq:goulianos}), the suppression factor $r_{{\rm h}}$ is assumed to be
$x_{\Pomeron}$-independent.

Next we discuss the hard coherent dijet production in proton-nucleus
scattering via the e.m.~mechanism, when
the nucleus coherently emits a quasi-real photon which interacts with the proton
and diffractively produces two hard jets,
$p+A  \to p+\gamma+A \to 2\,{\rm jets}+X+A$,
see Fig.~\ref{fig:ultraperipheral}.
The 
corresponding cross section can be written as a sum of the
resolved and direct photon contributions (the separation into the 
resolved and direct components is only meaningful in the leading-order
calculation)
\begin{eqnarray}
\frac{d^3 \sigma^{p+A \to 2\,{\rm jets}+X+A}_{{\rm e.m.}}}{d x_1 d p_T^2 d x_{\Pomeron}}  &\propto& r_{{\rm e.m.}}\sum_{\substack{i,j,\\k,l=q,\bar{q},g}} f_{i/p}(x_1,Q_{{\rm eff}}^2)\frac{n(x_{\Pomeron})}{x_{\Pomeron}}
f_{j/\gamma}(\beta,Q_{{\rm eff}}^2) \overline{\sum}|{\cal M}(ij \to kl)|^2 \frac{1}{1+\delta_{kl}} \nonumber\\
& +& \sum_{\substack{i,j,\\k,l=q,\bar{q},g}} f_{i/p}(x_1,Q_{{\rm eff}}^2)\frac{n(x_{\Pomeron})}{x_{\Pomeron}} \delta(\beta-1) \overline{\sum}|{\cal M}(i\gamma \to kl)|^2 \frac{1}{1+\delta_{kl}}
\,,
\label{eq:cs_em}
\end{eqnarray}
where $n(x_{\Pomeron})$ is the flux of equivalent photons \cite{Baur:2001jj} expressed 
in terms of $x_{\Pomeron}$ instead of the photon energy $\omega$ (note the
factor $1/x_{\Pomeron}$ coming from the $1/\omega$ in the spectrum of the
equivalent photons); 
$f_{j/\gamma}$ is the PDF of the real photon;  $\overline{\sum}|{\cal M}(i\gamma \to kl)|^2$
are invariant matrix elements for the direct photon-parton scattering,
 see Table 7.9 in
\cite{Ellis}; $r_{{\rm e.m.}}$ is a phenomenological factor describing the
factorization breaking for the resolved (hadron-like) component of
the real photon. The exact value of  $r_{{\rm e.m.}}$ is uncertain: It
ranges from $r_{{\rm e.m.}}=0.34$~\cite{Klasen:2004ct}
 to $r_{{\rm e.m.}} \approx 1$
with large errors~\cite{Chekanov:2001bw}. Since our analysis is a simple leading-order 
estimate, we conservatively take $r_{{\rm e.m.}}=0.5$.

The flux of equivalent photons approximately equals~\cite{Baur:2001jj}
\begin{equation}
n(x_{\Pomeron}) \approx \frac{2 Z^2 \alpha_{{\rm e.m.}}}{\pi} \ln \left( \frac{\gamma}{R_A x_{\Pomeron} \,p_{{\rm lab}}} \right) \,,
\label{eq:flux}
\end{equation}
where $Z$ is the nuclear charge; $\gamma$ is the Lorentz factor 
($\gamma \approx 3000$ for
$p\,Pb$ scattering at the LHC~\cite{Morsch}); 
$R_A=1.145\,A^{1/3}$ fm is the effective nuclear  radius; 
$p_{{\rm lab}}$ is the momentum of the nucleus in the laboratory frame 
($p_{{\rm lab}} \approx 2.75$ TeV for
$p\,Pb$ scattering at the LHC).
In practice, we used a more precise formula for the flux of the equivalent photons, which
reduces the result of Eq.~(\ref{eq:flux}) by 11\%~\cite{Zhalov}.

We are now ready to estimate the ratio of the hard diffractive dijet cross sections
corresponding to the hard and e.m.~mechanisms,
$R$,
\begin{equation} 
R(\beta,x_{\Pomeron},p_T)=\frac{d^3 \sigma^{p+A \to 2\,{\rm jets}+X+A}}{d x_1\,d p_T^2\,d x_{\Pomeron}} \Bigg/
 \frac{d^3 \sigma^{p+A \to 2\,{\rm jets}+X+A}_{{\rm e.m.}}}{d x_1 \,d p_T^2 \,d x_{\Pomeron}} \,,
\label{eq:r}
\end{equation}
where the involved cross section are given by Eqs.~(\ref{eq:cs_hard}) and 
(\ref{eq:cs_em})
with the equal coefficients of proportionality.
In the simplified kinematics that we use, at given $p_T$ and $x_{\Pomeron}$, the ratio
$R$ depends only on $\beta$.

\begin{figure}[t]
\begin{center}
\epsfig{file=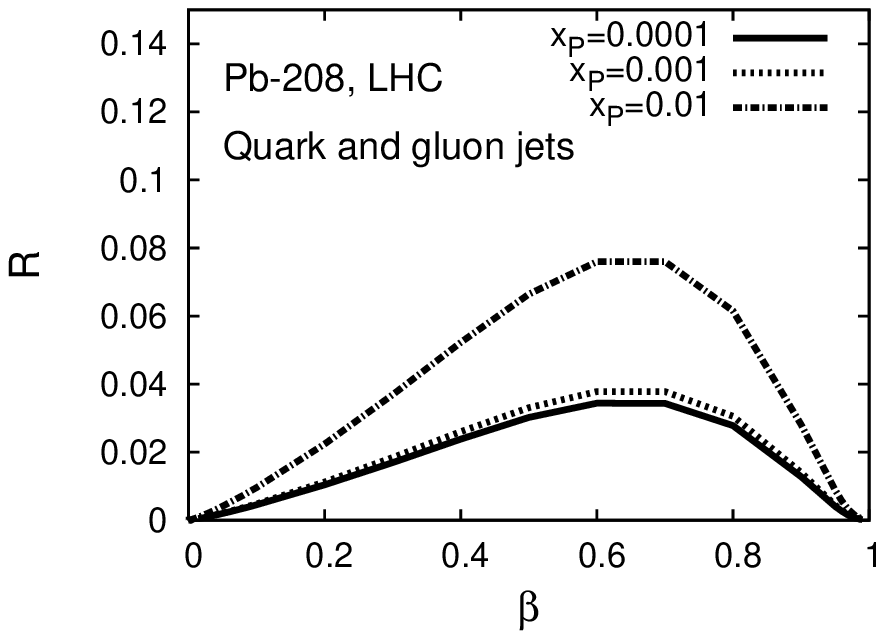,width=8cm,height=8cm}
\epsfig{file=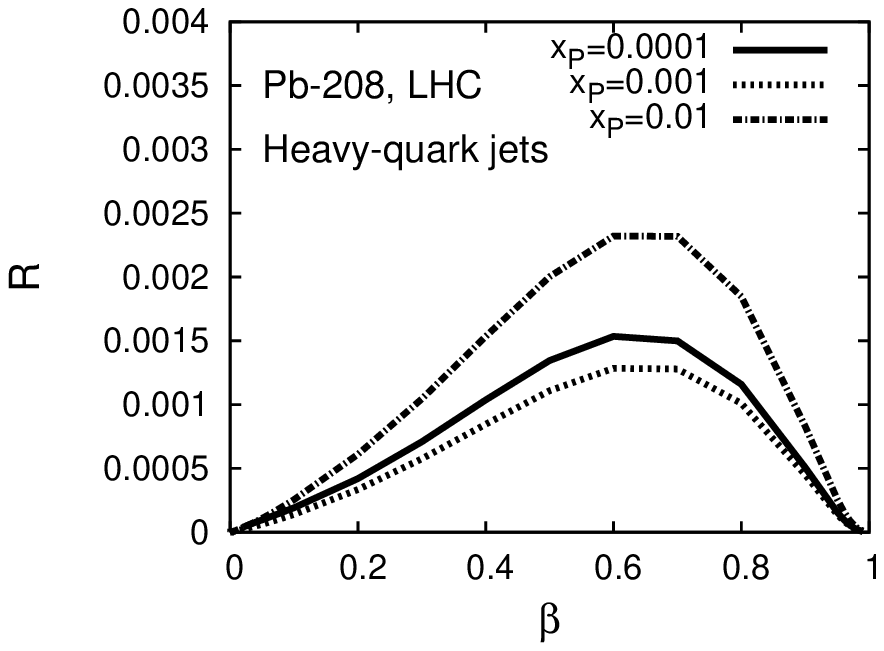,width=8cm,height=8cm}
\caption{The suppression of hard diffractive dijet (quark and
gluon jets) production compared to e.m.~coherent dijet production in
proton-Pb scattering at the LHC. The suppression factor $R$ of Eq.~(\ref{eq:r})
at $p_T=5$ GeV and $x_{\Pomeron}=10^{-4}$, $10^{-3}$ and $10^{-2}$ 
as a function of $\beta$. The left panel corresponds to quark and gluon jets; the right panel corresponds to heavy-quark jets.
}
\label{fig:jets_all}
\end{center}
\end{figure}

We considered two cases: The dijet production summed over gluon and quark jets
and the production of two heavy-quark jets ($c$ and $b$ quarks). The resulting ratios $R$ at $p_T=5$ GeV and 
$x_{\Pomeron}=10^{-4}$, $10^{-3}$ and $10^{-2}$ as functions of $\beta$ are presented
in Fig.~\ref{fig:jets_all}.
The left panel corresponds to quark and gluon jets; the right panel
corresponds to heavy-quark jets.

The results presented in Fig.~\ref{fig:jets_all} 
deserve a detailed discussion. The dependence of the ratio $R$ on 
$x_{\Pomeron}$ is not too strong and
can be explained as follows.
The main  contribution to the 
$x_{\Pomeron}$-dependence of $R$ at fixed $\beta$ comes from the changing 
of $x_1$.
 As 
$x_{\Pomeron}$ is decreased, $x_1$ is increased, which diminishes the role
played by the gluons in the projectile. As explained in the following, it
is the gluon contribution that increases $R$. Hence, $R$ decreases with
decreasing $x_{\Pomeron}$. 
Note that 
the dependence of diffractive PDFs on $x_{\Pomeron}$,
$f_{j/N}^{D(3)}(\beta,x_{\Pomeron},Q_{{\rm eff}}^2) 
\propto 1/x_{\Pomeron}^{1+2\,\epsilon}$, see Eq.~(\ref{eq:pomeron_flux}), 
is similar to the $1/ x_{\Pomeron} \ln(1/x_{\Pomeron})$-behavior
 of the e.m.~cross section. Therefore, these two factors 
weakly affect the $x_{\Pomeron}$-dependence of
$R$.

The dependence of $R$ on $\beta$ is much faster and reflects different
shapes of the proton diffractive PDFs and PDFs of the real
photon. While the proton diffractive PDFs times $\beta$ are flat in the $\beta \to 0$
limit, the photon PDFs times $\beta$ grow. This explains why $R$ approaches
zero when $\beta$ is small. In the opposite limit, $\beta \to 1$,
diffractive PDFs vanish and the e.m.~contribution wins over due to the 
non-vanishing
direct photon contribution: $R \to 0$ as $\beta \to 1$.

In Fig.~\ref{fig:jets_all}, the ratio $R$ at its peak is much larger
for the production of quark and gluon jets than for the production
of heavy-quark jets.
An examination shows that
 this effect is due to the large gluon diffractive 
PDF, which in tandem with
the large $gg \to gg$ hard parton invariant matrix element~\cite{Ellis},
 works to increase $R$ in the presence of the gluon jets.

We used the following input in 
our numerical analysis of the ratio $R$. We used the LO parameterization
of the real photon PDFs from Ref.~\cite{Gluck:1991jc}. We have also checked that the use
of a different parameterization~\cite{Abramowicz:1991yb}
 leads to rather similar 
predictions. 

For the nucleon diffractive PDFs, we used the recent QCD fit to the H1 data on 
hard inclusive diffraction in DIS on hydrogen~\cite{unknown:2006hy,:2006hx}.
 The suppression factor  $\lambda^j$,
which enters Eq.~(\ref{eq:r}) at the scale $Q^2=Q_{{\rm eff}}^2=4\,p_T^2=100$ GeV$^2$,
was evaluated using Eq.~(\ref{eq:lambda_parton}) with $\sigma_{{\rm eff}}^j(x,Q^2)$ at the same $Q^2=Q_{{\rm eff}}^2=100$ GeV$^2$ scale, see 
Eq.~(\ref{eq:sigmaeff}).

The $\delta$-function for the direct photon contribution was numerically modeled in
 the following simple form
\begin{equation}
\delta(\beta-1)=\frac{1}{\pi} \frac{\epsilon}{(\beta-1)^2+\epsilon^2} \,, \quad {\rm with} \quad
 \epsilon=0.01 \,.
\end{equation}

\begin{figure}[t]
\epsfig{file=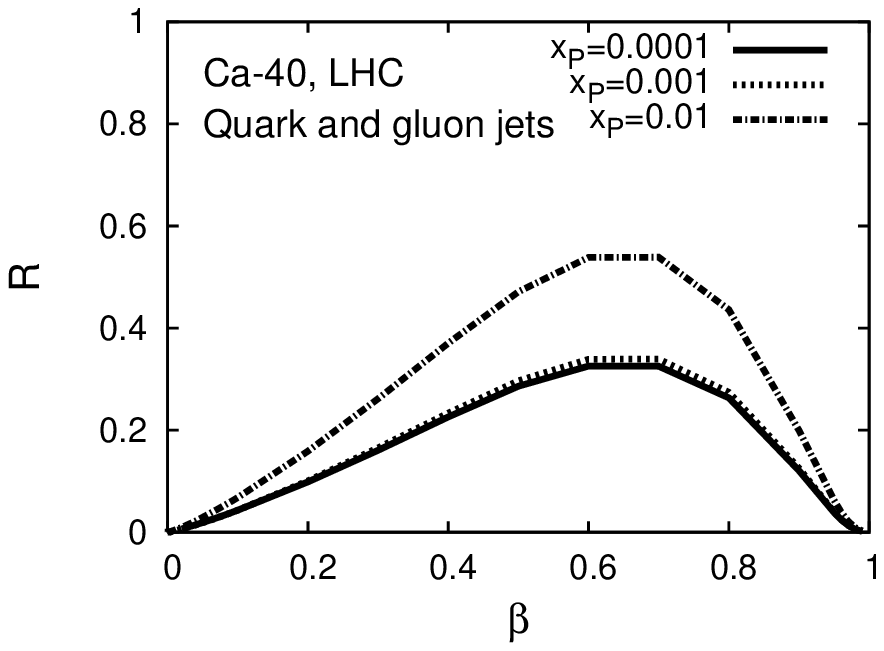,width=8cm,height=8cm} 
\epsfig{file=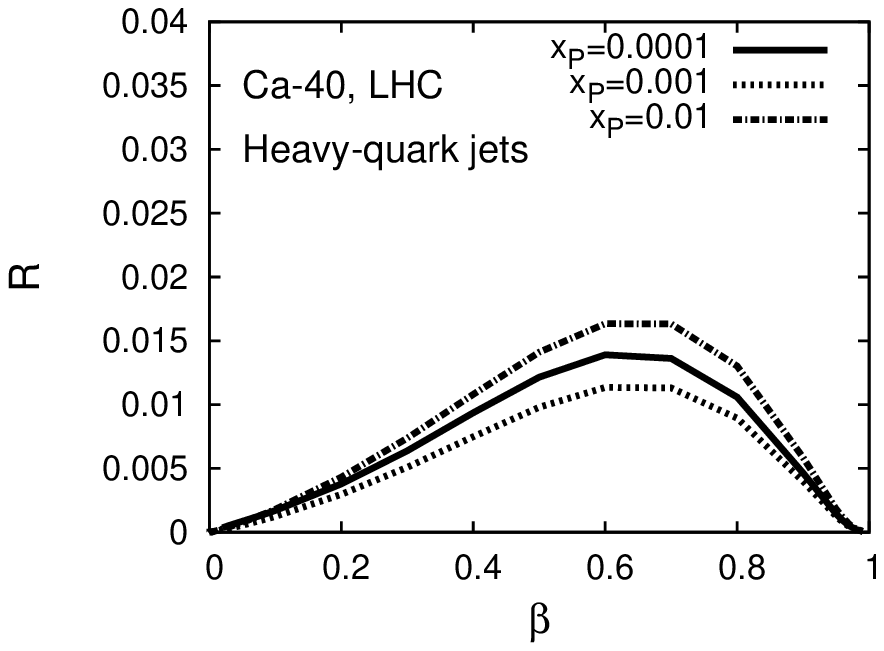,width=8cm,height=8cm}
\caption{The suppression factor $R$ of Eq.~(\ref{eq:r}) for proton-$^{40}$Ca scattering
at the LHC.
The labeling of the curves is the same as in Fig.~\ref{fig:jets_all}.
}
\label{fig:jets_ca40}
\end{figure}

It is instructive to examine how our predictions for the suppression factor $R$ change,
when the heavy nucleus of $^{208}$Pb is replaced by a lighter nucleus of $^{40}$Ca.
Note that for $p\,Ca$ scattering at the LHC, $\sqrt{s}=9.9$ TeV and 
$\gamma \approx 3700$~\cite{Morsch}.
 We expect that
the ratio $R$ will significantly increase because of the reduction of the flux of
the equivalent photons [the flux is proportional $Z^2$~(\ref{eq:flux})].

Figure~\ref{fig:jets_ca40} presents the ratio $R$ for
$^{40}$Ca. The labeling of the curves is the same as in Fig.~\ref{fig:jets_all}.
 As can be seen from the comparison of  Figs.~\ref{fig:jets_ca40}
and \ref{fig:jets_all}, the ratio $R$ increases by the factor
$\approx 7-10$, when going from $^{208}$Pb to $^{40}$Ca.

\begin{figure}[t]
\epsfig{file=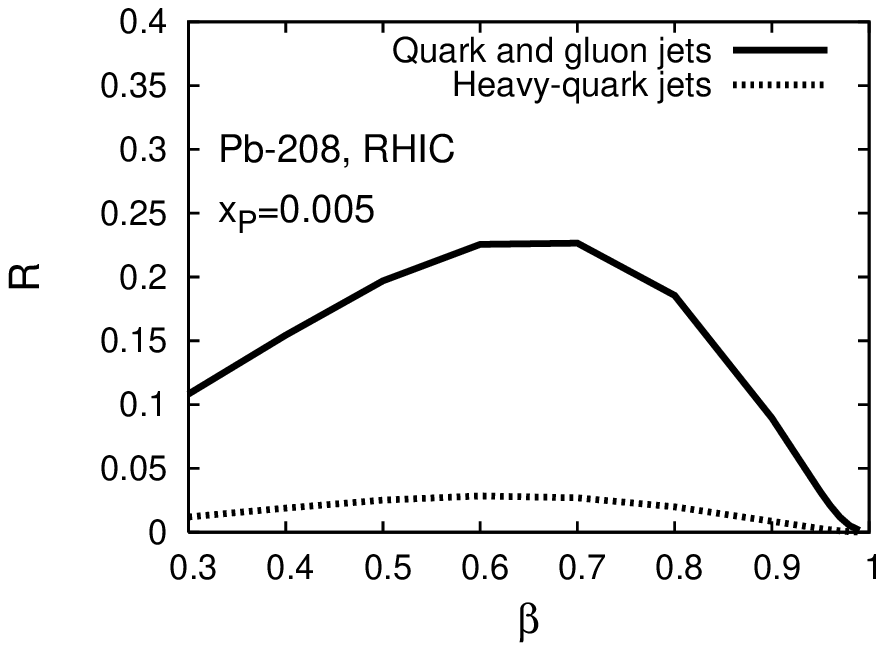,width=8cm,height=8cm} 
\epsfig{file=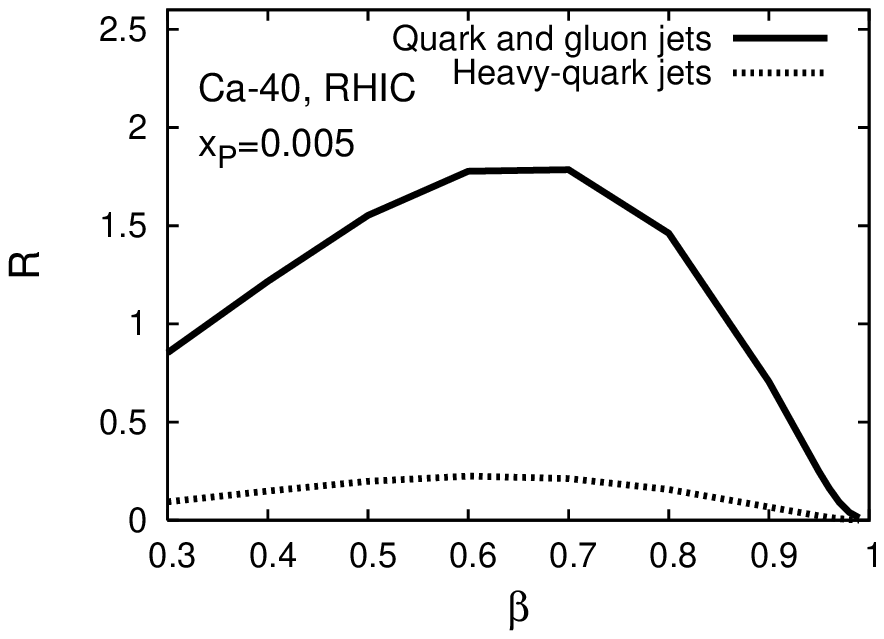,width=8cm,height=8cm}
\caption{The suppression factor $R$ of Eq.~(\ref{eq:r}) in the RHIC
kinematics and at  $p_T=5$ GeV and $x_{\Pomeron}=5 \times 10^{-3}$ 
as a function of $\beta$. The solid curves correspond to 
quark and gluon jets; the dotted curves correspond to heavy-quark jets.}
\label{fig:rhic}
\end{figure}

Besides the LHC, RHIC also has a potential to measure hard diffraction in
proton-nucleus scattering.
We consider a typical example of the corresponding 
RHIC kinematics with 250 GeV protons 
scattering on 100 GeV/per nucleon nuclei 
(the corresponding $\sqrt{s} \approx 320$ GeV and the
Lorentz dilation factor is $\gamma \approx 100$).
Producing sufficiently high 
diffractive masses, e.g. $M_X^2=500$ GeV$^2$, one accesses the 
typical kinematics of 
hard diffraction, $x_{\Pomeron}=5 \times 10^{-3}$ and $\beta > 0.3$.
Note also that the suppression of hard diffraction at RHIC is approximately 
four times smaller
than at the LHC, see Eq.~(\ref{eq:goulianos3}).

We studied the suppression factor $R$ of Eq.~(\ref{eq:r}) in the considered 
RHIC kinematics at $p_T=5$ GeV. The resulting values of $R$ as a function of
$\beta$ are presented in Fig.~\ref{fig:rhic}.
The solid curves correspond to 
quark and gluon jets; the dotted curves correspond to heavy-quark jets.

As seen from Fig.~\ref{fig:rhic}, the factor $R$ at RHIC is larger than 
at the LHC. This is mostly a consequence of the decrease
 of the flux of equivalent photons when going from the LHC to the RHIC 
kinematics.

\section{Conclusions}

Using the Glauber-Gribov multiple scattering formalism and
the
leading twist theory of nuclear shadowing, we developed a
 method for the calculation of  coherent
 hard diffraction processes off nuclei. We showed that soft multiple rescatterings
lead to the factorization breaking of hard diffraction in proton-nucleus  
scattering, which is larger than the well-known factorization breaking 
of diffraction in hadron-nucleon scattering.

Based on these results, we compare the hard diffractive to e.m.~mechanisms of
hard coherent production of two jets in proton-nucleus scattering.
We study the $x_{\Pomeron}$, $\beta$ and $A$-dependence
of the ratio of the dijet production cross sections due to the two effects,
 $R$, at the LHC and RHIC kinematics. We
separately study the case when the final jets consist of quarks and gluons and the
case when the final jets consist of heavy ($c$ and $b$) quarks.

Our results can be summarized as follows.
For proton-$^{208}$Pb scattering at the LHC, hard diffraction is suppressed compared to
the e.m.~contribution, especially at $x_{\Pomeron}=10^{-4}$ and large $\beta$, e.g.
$\beta > 0.8$, see Fig.~\ref{fig:jets_all}.
 The suppression is very strong for the production of heavy-quark
jets, see the right panel of Fig.~\ref{fig:jets_all}.
The physical reason of the suppression is the strong coherent Coulomb field of 
$^{208}$Pb, which enhances the e.m.~mechanism of hard diffraction.

Replacing $^{208}$Pb by $^{40}$Ca, the hard diffractive mechanism becomes compatible to
the e.m.~one in the case of the production of quark and gluons jets, see
the left panel of Fig.~\ref{fig:jets_ca40}.
However, like in the case of $^{208}$Pb, the production of heavy-quark jets is
dominated by the e.m.~mechanism, see
the right panel of Fig.~\ref{fig:jets_ca40}.

As a result of the smaller Lorentz dilation factor $\gamma$ at RHIC,
 the factor $R$ at the
 RHIC kinematics is larger than at the LHC.

Our results suggest the following experimental strategies. First, the use of heavy nuclei
 in $p\,A$ scattering at the LHC will provide a clean method to study  hard
 real photon-proton scattering at the energies exceeding the HERA energies by the
 factor of ten. Second, taking lighter nuclei and choosing the appropriate kinematics,
where the e.m.~contribution 
can be controlled, one can effectively study the factorization
breaking in nuclear diffractive PDFs. Third, in the same kinematics, a comparison
of the dijet diffractive production to the heavy-quark-jet diffractive production
will measure the nuclear screened diffractive gluon PDF. It  can be compared to the nuclear
diffractive PDFs, which will be measured in nucleus-nucleus ultraperipheral
collisions at the LHC.

\acknowledgements

We would like to thank 
K.~Goulianos for  valuable discussions of the factorization breaking
in $p\,{\bar p}$ diffraction
and M.~Zhalov for the discussions of the suppression 
factor for hard diffraction
and the calculation of the correction to the flux of equivalent photons.  
This
work is supported by the Sofia Kovalevskaya Program of the Alexander
von Humboldt Foundation (Germany) and DOE (USA).
M.S.~thanks the Frankfurt Institute for Advanced Studies at Frankfurt
University for the hospitality during the time when this work was 
completed.

\end{document}